\documentclass[twocolumn,twoside,prd,floatfix,letterpaper]{revtex4}

\usepackage[pdftex]{graphicx} 
\usepackage{color}
\usepackage{amsfonts}
\usepackage{tfrupee}
\usepackage[colorlinks,hyperindex]{hyperref}
\usepackage[all]{hypcap}

\def \lastDataDate {December 31, 2021}
\newcommand{\ldds}[1]{#1}  

\def \FigRandomWalk {Figure~\ref{fig:RandomWalk}}
\def \FigMain {Figure~\ref{fig:Indices}}
\def \FigIndices {Figure~\ref{fig:Indices}}
\def \FigUSCompanies {Figure~\ref{fig:USCompanies}}
\def \FigChinaCompanies {Figure~\ref{fig:ChinaCompanies}}
\def \FigsWorldUSChina {Figures~\ref{fig:Indices}, \ref{fig:USCompanies}, and \ref{fig:ChinaCompanies}}
\def \FigSEC {Figure~\ref{fig:SECGameStopReport}}
\def \FigMeme {Figure~\ref{fig:MemeStocks}}

\def \citeKnuteson {\cite{knuteson2016,knuteson2018,knuteson2019,knuteson2020,knuteson2021}}
\def \theExplanation {\cite{knuteson2016,knuteson2018,knuteson2019}}
\def \RefsTheExplanation {Refs.~\theExplanation}
\def \citeAllOvernightIntradayLiterature {\cite{knuteson2016,knuteson2018,knuteson2019,knuteson2020,knuteson2021,cooper2008return,lachance2015night,qiao2020overnight,kelly2011returns,berkman2012paying,branch2012overnight,lou2019tug,bogousslavsky2021cross,lachance2020etfs}}

\def \M {\ensuremath{M}}
\def \spSOO {S\&P~500}

\begin{document}

\title{They Still Haven't Told You}
\author{Bruce Knuteson}
\noaffiliation

\begin{abstract}
The world's stock markets display a decades-long pattern of overnight and intraday returns seemingly consistent with only one explanation:  one or more large, long-lived quant firms tending to expand its portfolio early in the day (when its trading moves prices more) and contract its portfolio later in the day (when its trading moves prices less), losing money on its daily round-trip trades to create mark-to-market gains on its large existing book.  In the fourteen years since this extraordinary pattern of overnight and intraday returns was first noted in the literature, no plausible alternative explanation has been advanced.  The main question remaining is therefore which of the few firms capable of profitably trading in this manner are guilty of having done so.  If any of this is news to you, it is because the people you trust to alert you to such problems still haven't told you.
\end{abstract}

\maketitle

\vspace{-0.1in}

\tableofcontents

\section{Facts}

As in our previous articles on this topic~\citeKnuteson, the basic facts of interest are a remarkable pattern of overnight and intraday returns in the world's stock markets over the past three decades.  (An ``intraday return'' is the return from market open to market close.  An ``overnight return'' is the return from market close to the next day's market open.)  This remarkable pattern is robust, well established~\citeAllOvernightIntradayLiterature, undisputed, and easily reproducible.  You can recreate all of the plots in this article using publicly available data~\cite{yahooFinance} and our publicly available code~\cite{thisArticleWebpage}.

\begin{figure*}[tp]
\includegraphics[width=7in]{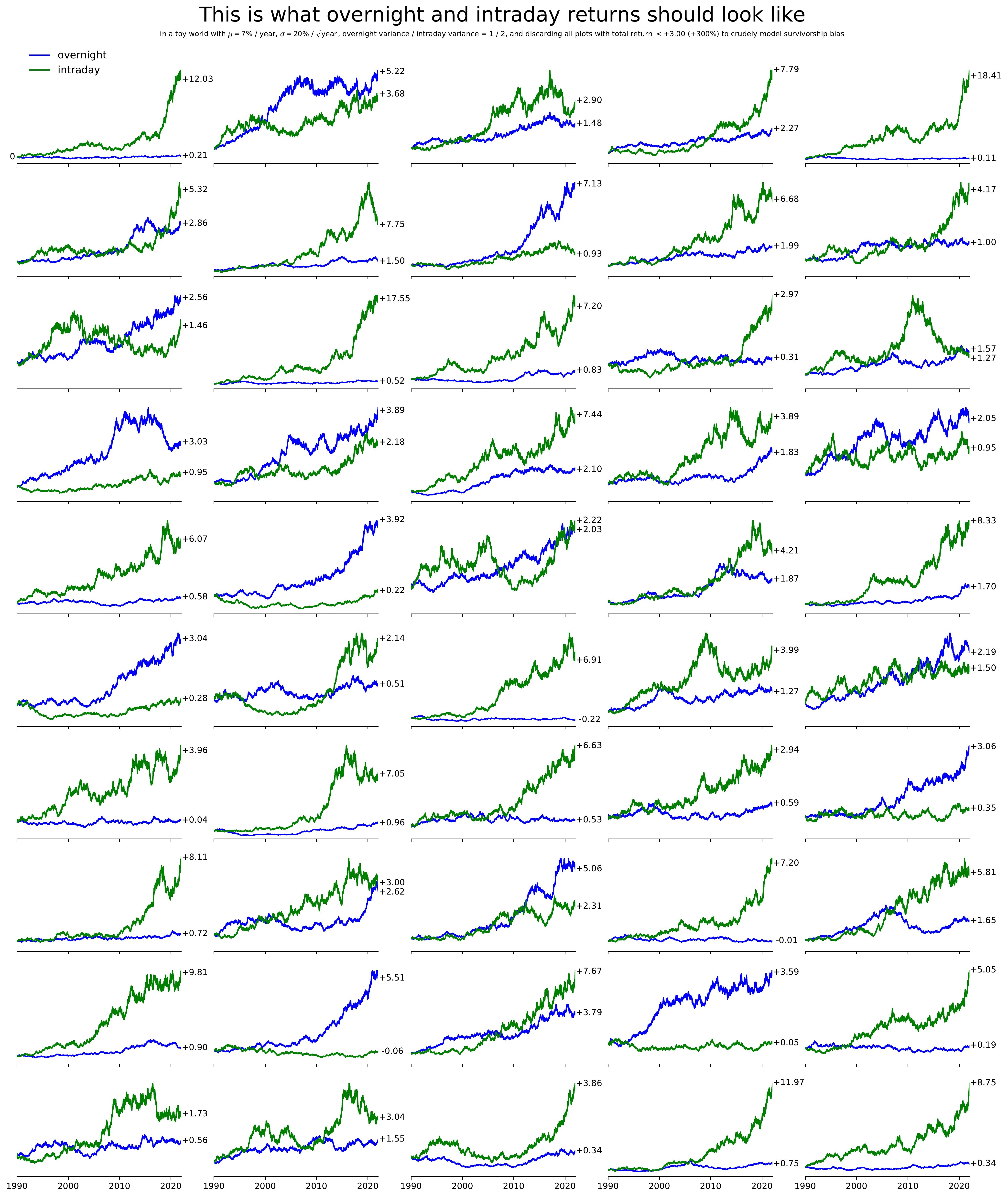}
\caption{\label{fig:RandomWalk}Fifty examples (generated with random numbers) of what plots of overnight and intraday returns should look like.  The horizontal axis spans January 1, 1990 to \lastDataDate.  The (linear) vertical axis of each plot extends from $-1$ ($-100\%$) (bottom of plot) through $0$ (where the blue and green curves start, at left) to the largest cumulative return achieved (top of plot).  The green curve shows cumulative intraday returns (from market open to market close).  The blue curve shows cumulative overnight returns (from market close to the next day's market open).  Thus, for example, if you had invested \$1 thirty-two years ago in the top left plot and had gotten only intraday returns, you would have netted a profit of \ldds{$+\$12.03$} (and realized a return on your original \$1 investment of \ldds{$+1{,}203\%$}).  The code used to make this figure is available at Ref.~\cite{thisArticleWebpage}.}
\end{figure*}
\begin{figure*}[tp]
\includegraphics[width=7in]{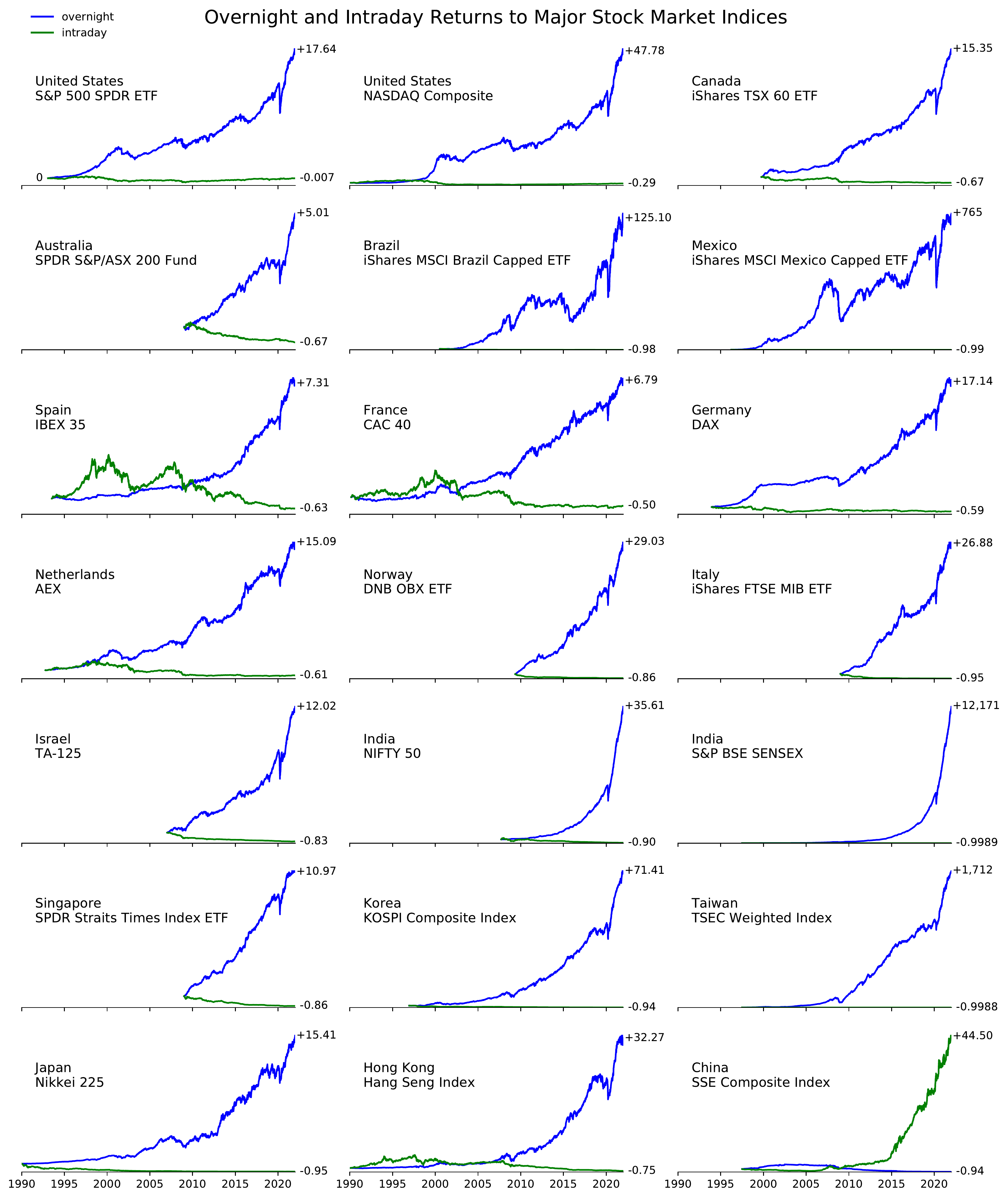}
\caption{\label{fig:Indices}Cumulative overnight (blue) and intraday (green) returns to twenty-one major stock market indices over thirty-two years, from January 1, 1990 to \lastDataDate.  The format of each plot is the same as \FigRandomWalk.  A return of $+1$ is equivalent to a return of $+100\%$.  If you had invested \rupee 1 in India's SENSEX on July 1, 1997 and had gotten only overnight returns, by \lastDataDate\ you would have made \rupee \ldds{12{,}171}, for a cumulative return of roughly \ldds{$+1{,}217{,}100\%$}.  If you had gotten only intraday returns, you would have lost \rupee \ldds{0.9989}, suffering a cumulative return of \ldds{$-99.89\%$}.  Data and code are publicly available~\cite{yahooFinance,thisArticleWebpage,yahooFinanceIndices}.  A version of this figure with logarithmic vertical scale (and showing the UK's FTSE~100 instead of Spain's IBEX~35~\cite{yahooFinanceFtse100}) is provided in Ref.~\cite{knuteson2021}.}
\end{figure*}
\begin{figure*}[tp]
\includegraphics[width=7in]{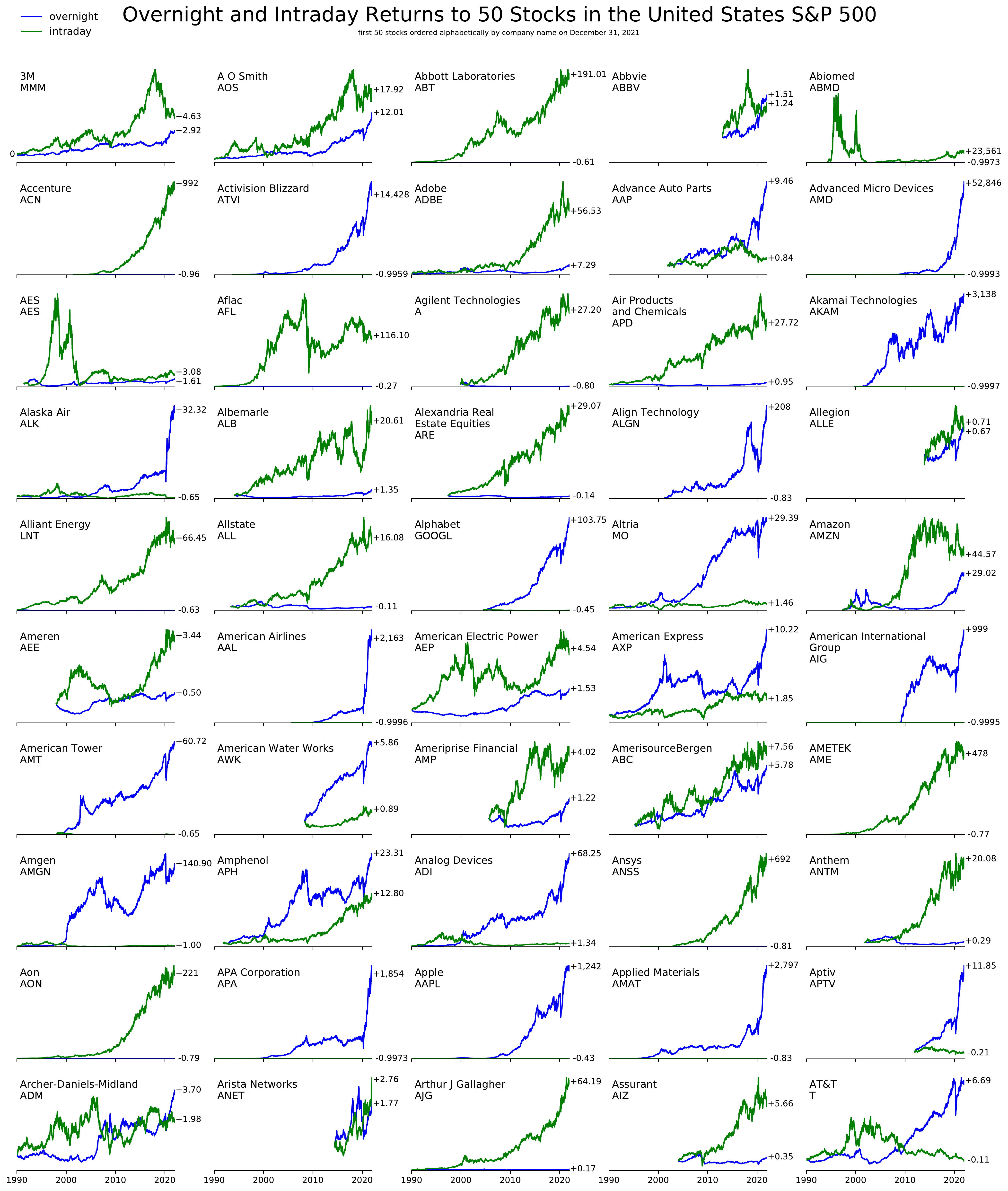}
\caption{\label{fig:USCompanies}Cumulative overnight (blue) and intraday (green) returns to the first fifty companies sorted alphabetically by name in the \spSOO\ index (as of \lastDataDate) from the start of 1990 (or the first date for which data are available) to \lastDataDate~\cite{SP500Constituents}.  The format of each plot is the same as \FigIndices.  If you had invested \$1 in Apple on January 1, 1990 and had gotten only overnight returns, thirty-two years later you would have made \$\ldds{$1{,}242$}, for a cumulative return of roughly \ldds{$+124{,}200\%$}.  If you had gotten only intraday returns, you would have lost \$\ldds{0.43}, suffering a cumulative return of \ldds{$-43\%$}.  Data and code are publicly available~\cite{yahooFinance,thisArticleWebpage}.  This figure is a small subset of the analysis in Ref.~\cite{lachance2015night}, presented as a picture.}
\end{figure*}
\begin{figure*}[tp]
\includegraphics[width=7in]{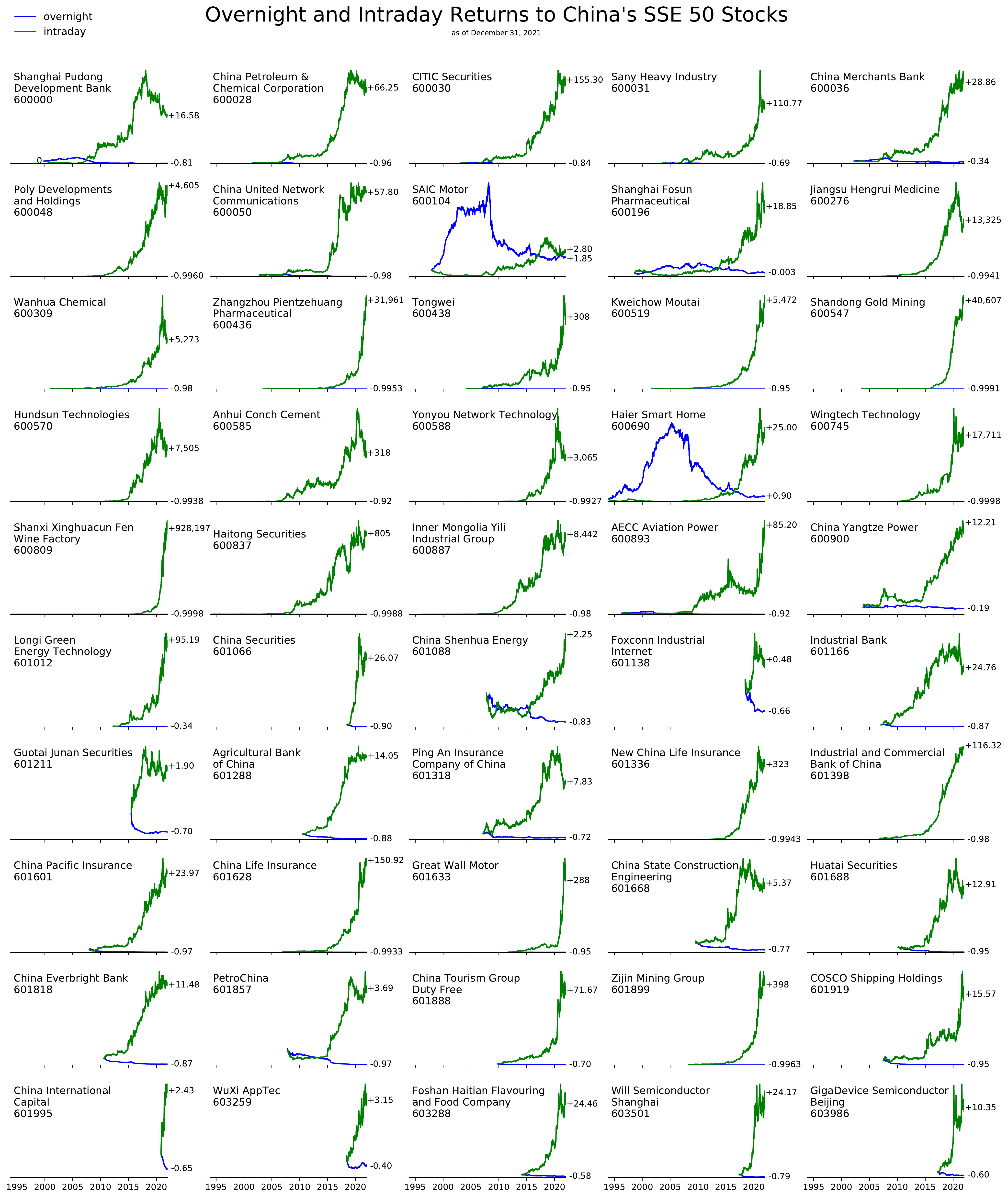}
\caption{\label{fig:ChinaCompanies}Cumulative overnight (blue) and intraday (green) returns to the fifty companies in China's SSE~50 index~\cite{ChinaSSE50Data}.  The format of each plot is the same as \FigUSCompanies.  If you had invested CN$\yen 1$ in China Life Insurance (601628) at the start of 2007 and had gotten only intraday returns, fifteen years later you would have made \ldds{CN$\yen 150.92$}, for a cumulative return of \ldds{$+15{,}092\%$}.  If you had gotten only overnight returns, you would have lost \ldds{CN$\yen 0.9933$}, suffering a cumulative return of \ldds{$-99.33\%$}.  Data and code are publicly available~\cite{yahooFinance,thisArticleWebpage}.}
\end{figure*}

To establish a baseline of normality relative to which the real world can be compared, let us think through what we expect a plot of a stock's overnight and intraday returns to look like.  In general, we expect a plot of stock returns to be a random walk with slight upward drift.  A positive return is expected, on average, for the bearing of risk.  If you compare the distribution of overnight returns with the distribution of intraday returns (as we do in Figure~4 of Ref.~\cite{knuteson2021}) for any of the individual stocks or indices in this article, you find the distribution of intraday returns is wider than the distribution of overnight returns.  Since returns are due to the bearing of risk and the intraday risk is greater than the overnight risk, you expect a larger intraday return than overnight return, on average.

Following this logic and generating some random numbers to visualize what it implies, \FigRandomWalk\ shows what plots of overnight and intraday returns from 1990 to present should look like~\footnote{\FigRandomWalk, generated with random numbers using the code at Ref.~\cite{thisArticleWebpage}, corresponds to a toy world in which markets are open five days every week (so the number of trading days is $n=261$ days/year), the expected return is $\mu=7\%$/year, the volatility is $\sigma=20\%/\sqrt{\rm{year}}$, one-third of each day's price variance realizes overnight (which is typical in the world's stock markets), and expected returns are due to the bearing of risk.  The expected return during a single overnight period is therefore $\mu_{\rm{o}}=\mu/3n$, the expected return during a single intraday period is $\mu_{\rm{i}}=2\mu/3n$, the overnight variance is $\sigma_{\rm{o}}^2=\sigma^2/3n$, and the intraday variance is $\sigma_{\rm{i}}^2=2\sigma^2/3n$.  To make each plot in \FigRandomWalk, we start on the first day, generating a random number from a Gaussian distribution with mean $\mu_{\rm{o}}$ and standard deviation $\sigma_{\rm{o}}$ for the overnight return and a random number from a Gaussian distribution with mean $\mu_{\rm{i}}$ and standard deviation $\sigma_{\rm{i}}$ for the intraday return.  We keep doing this for the thirty-two years between the start of 1990 and the end of 2021, showing the cumulative overnight returns in blue and the cumulative intraday returns in green.  Each of these fifty plots is constructed using exactly the same procedure, but different random numbers.  To crudely model survivorship bias (later figures do not show companies like Lehman Brothers, which have not survived), we throw away all plots in which the total return at the end of 2021 is less than $+300\%$.  The point is not any of these details, all of which are arguable and none of which really matter.  (Our choice of $\sigma$ is too large for some major indices and too small for most individual stocks, we ignore fat tails in the distribution of returns, our fifty independent plots do not reflect correlation among stocks or the effect of major historical events, and so on and so forth.)  The point of \FigRandomWalk\ is simply to show what you might reasonably expect plots of overnight and intraday returns to look like, providing a baseline against which later figures can be compared.}~\footnote{\FigRandomWalk\ is admittedly remedial -- a full page devoted to showing what a random walk with drift looks like -- but then at some level this entire topic is remedial.  Anyone who needs more than one article pointing out the existence of strikingly suspicious return patterns in financial markets to understand that there is a problem has something seriously wrong with him, and this is our sixth~\citeKnuteson.}.  The blue (overnight) and green (intraday) curves in \FigRandomWalk\ each look like a random walk.  They generally go up.  This is partly due to the previously mentioned positive expected return, and partly due to survivorship bias.  (In later figures, we will be showing companies of interest today, not companies that were of interest in 1990.  We have crudely modeled this survivorship bias in \FigRandomWalk\ by discarding plots with low total return.)  The green curve usually ends up higher than the blue curve.  \FigRandomWalk\ is sensible and understandable.  \FigRandomWalk\ is what you expect.

\FigIndices\ shows plots of overnight and intraday returns for twenty-one major stock market indices around the world.  Turn the page and compare \FigRandomWalk\ with \FigIndices.  See if you can tell a difference.

\FigUSCompanies\ shows plots of overnight and intraday returns for fifty stocks in the \spSOO\ index in the United States.  Take a moment, take a look, and take in how \FigUSCompanies\ looks nothing at all like Figures~\ref{fig:RandomWalk} and \ref{fig:Indices}.  Pay attention to the absurd numbers, too~\footnote{Our previous articles~\citeKnuteson\ provided cumulative returns as percentages.  Space constraints in Figures~\ref{fig:USCompanies} and \ref{fig:ChinaCompanies} make it difficult to express some of the most absurd numbers as percentages, so this article expresses all cumulative returns in Figures~\ref{fig:RandomWalk}, \ref{fig:Indices}, \ref{fig:USCompanies}, and \ref{fig:ChinaCompanies} as a fraction of unity.  Thus, for example, instead of showing the cumulative overnight return to Advanced Micro Devices (AMD) in \FigUSCompanies\ as \ldds{$+5{,}284{,}593\%$}, we show it as \ldds{$+52{,}846$} -- equally ridiculous, but four characters shorter.}.

Finally, noting that China is the exception in \FigIndices\ and having seen the US companies in \FigUSCompanies, take a guess as to what plots of overnight and intraday returns for companies in China look like.  Then check out \FigChinaCompanies, which shows plots of overnight and intraday returns for the fifty stocks in China's SSE~50 index.  Give yourself a break if your guess missed the mark, and give yourself time to fully appreciate the raw awesomeness of some of those numbers~\footnote{Our favorite company in \FigChinaCompanies\ is the Shanxi Xinghuacun Fen Wine Factory, where the cumulative overnight return is \ldds{$-99.98\%$}, the cumulative intraday return is \ldds{$+92{,}819{,}731\%$}, and the numbers make total sense if you are absolutely smashed.}.

You may consider \FigMain\ (together with other plots of overnight and intraday returns, like those in Figures~\ref{fig:USCompanies} and \ref{fig:ChinaCompanies}) to be the basic facts of interest in this article and in our previous articles on this topic~\citeKnuteson.  These basic facts are well established~\citeAllOvernightIntradayLiterature.  They are robust to using data from other data providers, to using prices at times shortly after (rather than at) market open and shortly before (rather than at) market close, and to other basic robustness checks~\citeAllOvernightIntradayLiterature.  Our articles on this topic are unusual among academic research in that you do not need to take our word for anything we say because you yourself can easily fact check everything we say~\cite{yahooFinance,thisArticleWebpage}.

\section{Proof}

We have previously provided the only plausible explanation so far advanced for \FigMain~\theExplanation.  In the fourteen years since the top left plot in \FigMain\ was first noted in the literature~\cite{cooper2008return}, no plausible alternative explanation for \FigMain\ has been proposed~\cite{knuteson2020,knuteson2021}.  The explanation in \RefsTheExplanation\ is the only explanation anyone has come up with that fits the facts.

``Firm proof'' is a set of facts seemingly consistent with only one explanation.  \FigMain\ is a set of facts seemingly consistent with only one explanation.  \FigMain\ itself therefore constitutes firm proof of the story told in \RefsTheExplanation.  Anyone wishing to argue otherwise must point to facts seemingly inconsistent with that story (in the same way we have disproved other attempted explanations~\cite{knuteson2020}) or clearly articulate an alternative explanation that fits the facts.

As far as your money is concerned, the burden of proof is on your regulator to convince you \FigMain\ is not a problem, not on us to convince you it is.  \FigMain\ shows suspicious return patterns in your investments.  That should concern you.  When you see something suspicious in one of your investments, you should ask for an explanation.  If you do not get an explanation that makes sense, you should take your hard-earned money elsewhere.

\begin{figure*}[tp]
\fbox{\includegraphics[width=7in,trim={0.5in 1.0in 0.5in 1.25in},clip]{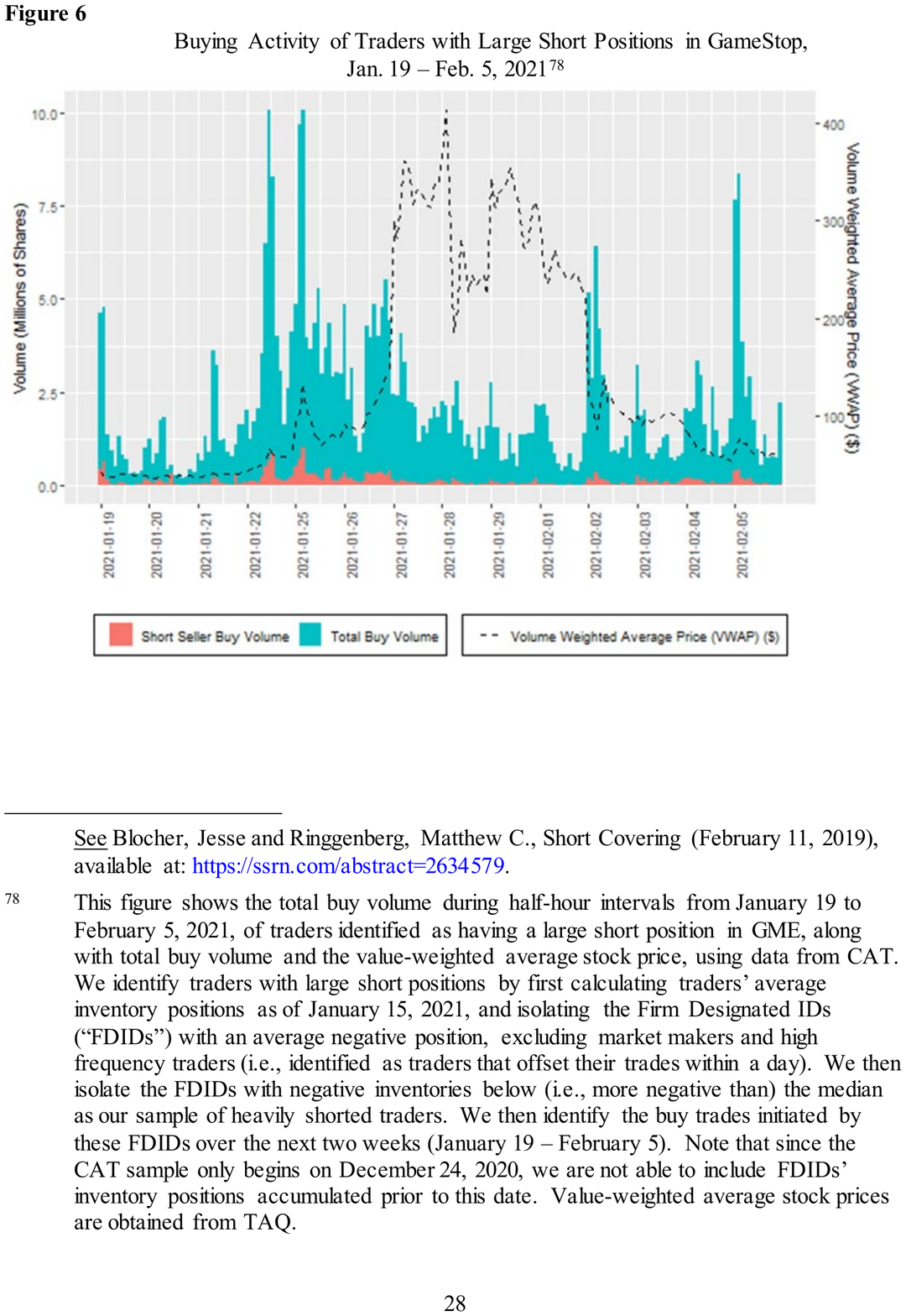}}
\caption{\label{fig:SECGameStopReport}Page 28 of the SEC's GameStop report~\cite{SECGameStopReport}, reproduced verbatim.  The image quality of the plot and labeling text is the same as Ref.~\cite{SECGameStopReport}.  The first and last sentences of footnote~78 incorrectly refer to ``value-weighted average stock price,'' a phrase that does not make logical sense and isn't a thing.  The analysis described on this page (and particularly the second to last sentence in footnote~78) undermines the popularly understood conclusion of Ref.~\cite{SECGameStopReport}, as we explain in Section~\ref{sec:Regulators}.}
\end{figure*}

\section{Regulators\label{sec:Regulators}}

If you trust your regulator to have conclusively investigated \FigMain, a comparison with Bernie Madoff may be useful.  (Feel free to replace Madoff -- the former chairman of the NASDAQ stock exchange who ran the largest Ponzi scheme in history~\cite{markopolos2010no}~\footnote{To reiterate:  the current Ponzi scheme world record holder was the chairman of a major United States stock exchange.} -- with a scandal closer to home, like Wirecard~\cite{WirecardFT}, where applicable.)  Everything about \FigMain\ is more problematic than Madoff.  The obvious suspects here are a few well established firms.  (With Madoff, there was just one.)  This issue is technically more complicated, since it involves actual trading.  (Madoff wasn't trading.)  Most importantly, no regulator wants to find it has missed an obvious case of market manipulation or fraud with broad scope.  (Madoff's scope was comparatively small.)  Any investigation of this issue will therefore involve a regulator who does not want to find evidence of market manipulation asking a well established quant firm to analyze its own trading for evidence of market manipulation -- a situation unlikely to produce evidence of market manipulation even in the presence of market manipulation.

If you think financial regulators are more competent now than they were at the times of their previous scandals, a recent report from the United States Securities and Exchange Commission (SEC) on curious price movements in the stock of GameStop~\cite{SECGameStopReport} will cure you of that.

The main popularly understood conclusion of the SEC's report, to the extent there are any conclusions to speak of, is that the remarkable increase in the price of GameStop's stock in January and February 2021 was not due to large short sellers covering their positions~\cite{levine2021gamestop,orland2021,salzman2021,martin2021,bain2021,rainey2021gamestop,mcenery2021,rearick2021,ongweso2021,harty2021}.  The analysis supporting this conclusion, which conveniently falls on a single page (page 28) of their report~\cite{SECGameStopReport} that we reproduce verbatim in \FigSEC, is a fine example of the quality of analysis of which the SEC is currently capable.

The SEC's analysis in \FigSEC\ purports to show that large short sellers were a small fraction of the total buy volume during the weeks of interest, and therefore the covering of short positions by large short sellers was not primarily responsible for the increase in GameStop's price.  The SEC's report has no author list or contact information for correspondence~\cite{SECGameStopReport}, denying us the opportunity to reach out to confirm our reading (just in case we misinterpret), but that presumably means the report was carefully written and designed to stand on its own, so take a minute to fully absorb \FigSEC.

The appearance of the gibberish phrase ``value weighted average price'' (twice!) in footnote~78 is concerning.  No person familiar with volume weighted average price would make this mistake.  Footnote~78, which describes the heart of the analysis behind the report's main conclusion, appears to have been written by someone whose understanding of both value and volume stops at the letter v.

Fortunately, the other sentences in footnote~78 contain no words that start with v, so we should be able to take them at face volume.  Reread the second to last sentence in footnote~78 in \FigSEC\ and chew on it for a minute.

Footnote~78 (and specifically its penultimate sentence) says the SEC does not know who all was short GameStop's stock.  If you established a huge short position in GameStop on December 15, 2020 and did not trade GameStop for the next month, the SEC's analysis thinks you have no position in the stock because the SEC's analysis is ignorant of everything that happened before December 24, 2020.  The title of the SEC's plot should more accurately be ``buying activity of some traders with large short positions in GameStop,'' with a note clearly admitting they don't really know what ``some'' means and therefore their orange histogram should be bigger and they don't really know how much bigger.  Since the point of the plot is that there isn't much orange, the fact that there really should be more orange and the reader doesn't have any sense of how much more orange there should be sort of defeats the point of the plot.  Beginning the second to last sentence of footnote~78 with ``Note that'' -- as though reminding you of a minor caveat they have previously mentioned rather than telling you for the first time a detail that undermines their entire analysis -- comes across as particularly slimy.  Not providing the number of shares that ended up being the threshold for ``large'' does little to increase the feeling of transparency.

\begin{figure}[thb]
\includegraphics[width=3.5in]{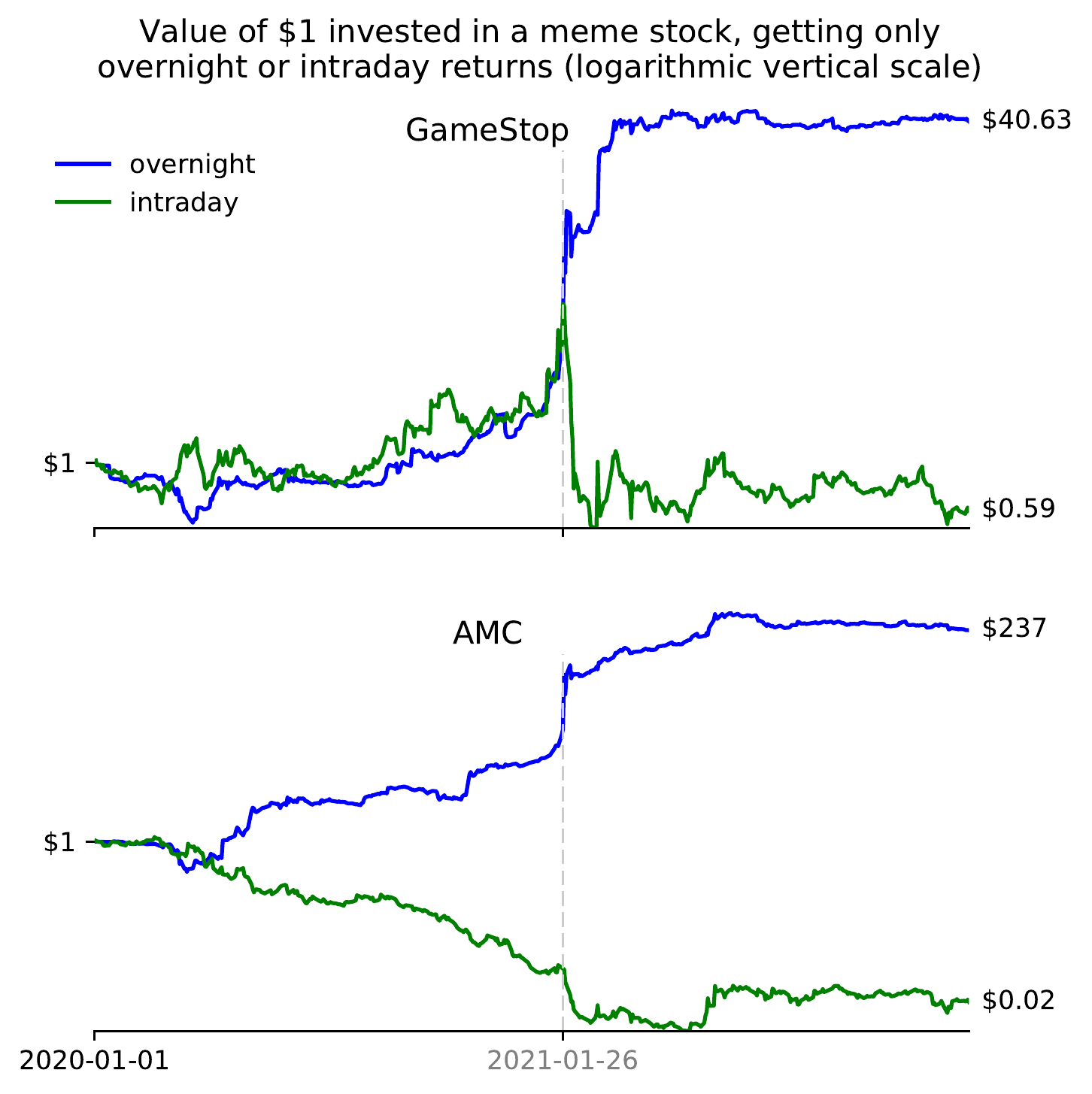}
\caption[Overnight/intraday returns to meme stocks]{\label{fig:MemeStocks}The value of \$1 of GameStop (top) and AMC (bottom) stock, invested at the start of 2020, getting only overnight or intraday returns.  The vertical axis of each plot has logarithmic scale.  One dollar invested in GameStop at the start of 2020, getting only intraday returns, would leave you with \ldds{\$0.59} on \lastDataDate, for a cumulative return of \ldds{$-41\%$}.  This figure first appeared in Ref.~\cite{knuteson2021}.}
\end{figure}

Having eviscerated the analysis purportedly supporting the SEC's main conclusion with a single paragraph, let us now review the clue we provided in Ref.~\cite{knuteson2021} (months before the release of the SEC's report) that contains more insight into the cause of the GameStop saga than the entirety of the SEC's report (which represents nine months of SEC staff investigation at the prompting of Congress).

\FigMeme\ shows overnight and intraday returns to GameStop's stock over the two years from January 1, 2020 to \lastDataDate, with logarithmic vertical scale.  You can easily make this plot yourself~\cite{yahooFinance,thisArticleWebpage}.  \FigMeme\ is striking:  all of the positive returns during the period of interest came overnight.  The SEC does not mention this remarkable fact anywhere in their report~\cite{SECGameStopReport}.  In fact, the SEC explicitly excludes overnight -- the period of greatest interest -- in their plot shown in our \FigSEC.  Although the SEC's plot does not explicitly label times of day, footnote~78 tells us each histogram bin spans thirty minutes, and there appear -- it is hard to tell because the image quality is so poor, almost as if the plot itself is embarrassed to be seen in the SEC's report and is actively trying to wriggle its way out of it -- to be thirteen bins per day, so the SEC's plot appears to show data from market open (9:30am) to market close (4pm).  Our \FigMeme\ shows that the period of greatest interest is from market close to market open.  Explicitly excluding the most important time period is exactly the sort of analysis choice you should expect from the SEC, and one that is fully consistent with the quality of the rest of the page shown in \FigSEC.

To be clear, we do not claim to know (or care in the slightest) whose trading caused the GameStop saga.  We showed \FigMeme\ in Ref.~\cite{knuteson2021} in an attempt to draw people's attention from a curiosity that does not matter (GameStop) to its overnight/intraday return pattern (top of \FigMeme) to the much more striking overnight/intraday return pattern in its companion meme stock AMC (bottom of \FigMeme) to the overnight/intraday return pattern in the world's major stock market indices (\FigMain), a problem that is tens of trillions of dollars more important.  We reprise \FigMeme\ in this article only because the SEC's GameStop report~\cite{SECGameStopReport}, released months after Ref.~\cite{knuteson2021}, provides us with an opportunity to assess the SEC's ability to understand the cause of anything.

So, to recap, the SEC by its own admission did not figure out whose trading caused the increase in GameStop's price~\footnote{\FigMeme\ disproves the SEC's claim the GameStop saga was due to ``positive sentiment'' -- unless the SEC means ``very positive sentiment while the market was closed and negative sentiment while the market was open,'' which is exactly as silly as it sounds.  This level of quality, from one of the world's premier financial regulators who spent nine months investigating with the attention of Congress, is an upper bound on the level of quality you can expect from your financial regulator on matters of less widespread and pressing interest.}, the SEC's analysis purportedly supporting its conclusion that the covering of short positions didn't cause the run up in price does not actually support that conclusion, the SEC ignored the overnight/intraday clue they might have focused on to understand the cause (\FigMeme), they did not mention (much less explain) the remarkable fact that all of the positive returns over the period of interest came overnight, at least one of the authors appears to be confused about value and volume, and the entire report generally reads more like a belated final write-up from an underwhelming summer intern than a definitive account from one of the world's premier financial regulators.

Our point is not to make fun of how many mistakes the SEC can pack into a single page consisting of nothing more than one plot and one footnote~\footnote{Ref.~\cite{knuteson2021} provided a gentle response to the argument that \FigMain\ must not be a problem because if it were, regulators would have addressed it.  Section~\ref{sec:Regulators} may be read as a more direct response to this argument.  Having to address this argument directly is unfortunate, since the substance of any such response (namely, that your regulator is indeed capable of turning a blind eye to \FigMain) is intrinsically and unavoidably offensive to people in a position to address \FigMain.  We try to make the point effectively but harmlessly, narrowly restricting the scope of most of our response to one page of one report from one regulator, and softening our response with innocuous trivialities, like the fourteen times ``2021'' appears unnecessarily in \FigSEC.}.  Our point is as serious as your money is to you.  If you had any illusion that the SEC is capable of determining whose trading is behind the extraordinary plots in Figures~\ref{fig:Indices} and \ref{fig:USCompanies}, Ref.~\cite{SECGameStopReport} should fully disabuse you of your fancy.  Ref.~\cite{SECGameStopReport} speaks values about the SEC's ability to understand the cause of things.  You don't want people who can't determine the cause of things policing anything related to your money.

The SEC's handling of the issue we are concerned about~\citeKnuteson\ is far worse.  Despite being fully aware of the problem~\cite{knuteson2021}, the SEC has publicly claimed that public equity markets in the United States function well, has encouraged the public to invest in them, and has not told the public about the suspicious returns in the United States stock market so obvious in many of the individual stocks in \FigUSCompanies\ and in the US indices in \FigIndices.  The SEC is telling you everything is fine while not telling you about suspicious features of your investments that clearly show everything is not fine.  The SEC is thus actively engaged in precisely the sort of behavior they are supposed to be protecting you against.

You definitely don't want people who have willfully and repeatedly chosen to not tell you about suspicious return patterns in your investments policing anything related to your money.

\section{Economists\label{sec:Economists}}

The failure of economists -- particularly those in academia -- to warn you about the problem so obvious in \FigMain\ is just as egregious as the failure of your regulators.  Key parts of their failure can be attributed directly to specific features of their incentive system.

The peer review process strongly favors articles with an explanatory narrative, disfavoring articles without such a narrative that clearly lay out puzzling empirical facts and clearly point out the puzzle.  Refs.~\cite{cooper2008return,lachance2015night}, which clearly lay out the puzzling empirical facts in \FigMain\ and clearly point out the puzzle, were never published.  The papers on this topic that have entered the peer-reviewed literature are therefore those with facts presented to support stories that make no sense, instead of those like Refs.~\cite{cooper2008return,lachance2015night} that most clearly lay out the facts and describe the puzzle~\footnote{You can see what we mean by comparing Ref.~\cite{lachance2015night} with Ref.~\cite{lachance2020etfs}, published five years and peer review later.  The insightful discussion of the pattern of overnight and intraday returns among individual stocks in Ref.~\cite{lachance2015night} (which we show as \FigUSCompanies) is nowhere to be found in Ref.~\cite{lachance2020etfs}, presumably in part because it does not fit the story told in Ref.~\cite{lachance2020etfs}.  We fault the peer review process and credit the author, who got it right the first time in Ref.~\cite{lachance2015night}.}.

Academics get no credit for seconding something someone else has already said.  Since we have clearly established ourselves as the ones who will receive full credit for the explanation in \RefsTheExplanation, no academic has any incentive whatsoever to make \FigMain\ and its implications more widely known.  Worse still, as time passes it becomes ever harder to acknowledge \FigMain\ as a problem while saving face~\footnote{Even worse still, anyone seeking to warn a wider audience of a problem missed by those closest to it must explain to the wider audience the failure of those closest to it.  Effectively doing this further warps the reputational incentives of (and increases the reputational stakes for) those closest to the problem.  We write Sections~\ref{sec:Regulators} and \ref{sec:Economists} only because we believe the potential benefit to the wider audience exceeds the lamentable cost of making it still harder for regulators and economists to correct their position while saving face.  (For our part, we strongly prefer to be wrong.  The truly scary implications of the problem we see in \FigMain\ far outweigh any small satisfaction of being right.)  Addressing a wider audience is most likely to be effective if the wider audience is predisposed to be receptive to the warning (which is not true in this case) or if the basic facts can be easily confirmed and the logic is straightforward (which is true in this case).}.

If you wanted to create an incentive system that appears fine to outsiders but is intentionally designed to keep people from recognizing \FigMain\ as a problem and raising the appropriate alarm, you would copy the current incentive system in academic economics.

Incentives aside, economists are poorly trained for this problem.  Economists generally believe markets are efficient, but nothing about \FigMain\ looks efficient.  Economists believe positive expected returns are due to the bearing of risk, but \FigMain\ shows negative intraday returns ranging from \ldds{-50\%} to \ldds{-99.89\%} outside the US and China.  Economists generally view markets as composed of a large number of small traders, but many small traders should not produce the striking consistency of the blue and green curves in \FigMain.  No general economic argument at the level of a stock market index is going to produce the striking differences in the overnight and intraday return patterns in individual stocks shown in \FigUSCompanies.  No economic argument at the level of individual stocks that accommodates \FigUSCompanies\ is going to produce the striking similarity in the overnight and intraday return patterns in the indices around the globe (excluding China) in \FigIndices.  The economist's view of the world hinders his recognition of \FigMain\ as a problem, and his limited toolkit contains nothing particularly helpful for addressing it~\footnote{In contrast, a physicist is inclined to view the market as a simple mechanical system.  Submitting an order perturbs the system.  The system then relaxes in a way that can be modeled and understood.  Her training provides the data analysis and theoretical tools needed to make progress.  She observes behavior across markets suggesting a simple, universal underlying model, further drawing her to this problem in a way others are not.}~\footnote{We use male pronouns where those at fault are overwhelmingly men.}.

Incentives and poor training aside, economists' failure to solve the puzzle of \FigMain\ can be equally attributed to poor taste in research topics.  Ignoring a plot like \FigMain\ is a jaw-dropping failure of judgment.  Separately, the puzzle of \FigMain\ is all about how individual orders affect prices, a topic that has been largely ignored by academic economists.  As we have previously noted~\cite{knuteson2018}, most of the academic research of value on this topic has been conducted by former physicists.  In our personal experience, also, former physicists are drawn to the topic of market impact in a way (and with a vigor) those from other backgrounds are not.  For all their talk about prices, economists have contributed surprisingly little of value to the understanding of how an order entered into the type of continuous-time, two-sided limit order book used in the world's stock markets quantitatively affects the price at and after that order is placed.

Again, you need not take our word for anything we say.  Ask an economist whose trading caused \FigMain.  If he provides more than one explanation, it means he doesn't really believe any of them -- if he really believed one of them was right, he wouldn't have bothered mentioning the others -- and you can avoid playing whack-a-mole by asking him to commit to the one explanation he considers most plausible.  Demand an article that fleshes out and pins down the explanation to ensure you both have a common understanding of what exactly the explanation entails and to avoid playing a version of whack-a-mole with one mole that shifts its shape whenever it gets whacked.  Read the article carefully and consider it seriously~\footnote{When you have finished reading the article with the alternative attempted explanation, you will probably know more about the article's contents than the referring economist does, because he probably has not recently carefully read it.}.  Decide for yourself whether the attempted explanation in the article explains the bizarre facts in \FigsWorldUSChina, including the large negative intraday returns in \FigMain, the striking consistency of the blue and green curves in \FigMain, the diverse overnight/intraday return patterns in individual stocks in \FigUSCompanies, and the communist-like homogeneity of the patterns in \FigChinaCompanies.

Our point is not to be critical for the sake of being critical~\footnote{Section~\ref{sec:Economists} may be read as a response (more direct than the one provided in Ref.~\cite{knuteson2021}) to the argument that \FigMain\ must not be a problem because if it were, economists would have raised an alarm.}.  Our point is every bit as serious as your money is to you.   Economists' sanguine attitude toward \FigMain\ is far less reassuring when you understand the mismatch between \FigMain\ and their actual expertise.  When listening to the experts, it is important to correctly identify the relevant expert for the problem at hand.

The correct explanation for \FigMain\ (and the correct theory of market impact) coming from a former physicist is not surprising or something economists should be embarrassed about.  Their failure to recognize \FigMain\ as a problem, their failure to bring \FigMain\ to your attention, and their failure to acknowledge and address the implications of the only explanation that fits the facts, on the other hand, are all shameful.  These failures have effectively ceded to us full credit for the explanation and all its many, varied, and highly consequential implications.  These include all the myriad (and mostly dire) implications of the culpable firms' trading having directly caused most of the suspiciously high returns to the world's stock markets over the past three decades.

You don't want people who can't recognize \FigMain\ as a problem making economic policy decisions that affect your money, and you certainly don't want people who knew about \FigMain\ and willfully and repeatedly chose to not tell you about suspicious return patterns in your investments anywhere near your money.

\section{Quants}

The explanation in \RefsTheExplanation\ begins by noting a generally known but widely underappreciated feature of the world's stock markets: early in the trading day, spreads are wide and depths are thin, while later in the trading day, spreads are narrow and depths are thick.  A trade early in the day therefore moves the price more than an equally sized trade later in the day.  Importantly, some of this price impact decays slowly, as described in Section~2 of Ref.~\cite{knuteson2019}.

A market participant with a sufficiently large portfolio can therefore expand his portfolio early in the day, when his trading moves prices more, and contract his portfolio later in the day, when his trading moves prices less, creating mark-to-market gains on his large existing book that exceed the cost of his daily round-trip trading.  We call this market manipulation the Strategy~\cite{knuteson2018,knuteson2019}, and we denote by \M~\cite{knuteson2016} any market participant using the Strategy.  \M's risk-return profile is most favorable if \M\ is generally market neutral, hedged against market returns (and possibly also other factor returns) over which \M\ has less short- and medium-term manipulative control.  To execute the Strategy, \M's trading must be systematic.  To benefit from the Strategy, \M's portfolio must be large.  \M\ is thus a large, generally market neutral quant firm using the Strategy.

To maximally benefit from the Strategy, \M\ expands his portfolio early in the day:  before, at, or shortly after market open~\footnote{For the purpose of this discussion (both for simplicity and because we consider it less important), we ignore trading after market close.}.  Because the impact of his earliest trading immediately starts decaying~\cite{knuteson2019} and because of the speed with which spreads narrow and depths thicken as the trading day begins in earnest, a wide range of trading profiles will lead to much of the price movement of \M's morning expansion appearing in the opening price~\footnote{How much of this price impact appears in the opening price certainly does depend on the profile of \M's trading -- if \M\ waits to start his trading until after market open, then none of it will appear in the opening price -- but as long as \M\ starts his trading before market open, the prediction that some meaningful fraction of the impact of \M's morning expansion will be reflected in the opening price is more robust than someone unfamiliar with the decay of market impact explained in Section~2 of Ref.~\cite{knuteson2019} and the fairly rapidly narrowing of spreads and thickening of depths as the trading day starts might expect.}.

\FigUSCompanies\ fits this picture even better than we have any right to expect given \M's slowly time-varying portfolio, the three decade time period, and the possibility of more than one \M\ contributing.  For the stocks \M\ likes to be consistently long (roughly one-third of those shown in \FigUSCompanies, consistent with the fraction noted in Ref.~\cite{lachance2015night}, and including ATVI, AMD, and AIG), you see an overnight/intraday return pattern consistent with \M\ expanding his long positions before and at market open, pushing the price up and causing the positive overnight (blue) curve, and then contracting his long positions later in the day, causing the negative intraday (green) curve.  For the stocks \M\ likes to be short (also roughly one-third of those shown in \FigUSCompanies, including ABT, ACN, and A), you see a negative overnight (blue) curve caused by \M\ pushing the price down as he expands his short positions before and at market open, and a positive intraday (green) curve caused by \M\ pushing the price up as he contracts his short positions later in the trading day.  The remaining one-third of the stocks in \FigUSCompanies\ show no clear pattern, corresponding to stocks \M\ was neither consistently long nor consistently short over this time.

This story~\theExplanation\ is easily falsifiable and highly predictive.  It predicts an \M\ using the Strategy.  This prediction is not something you or we can test with publicly available data, but it is a prediction a competent financial regulator can test~\footnote{\FigsWorldUSChina\ show a total of 121 plots, over one hundred of which don't look right.  These plots, which you can easily reproduce yourself~\cite{yahooFinance,thisArticleWebpage}, are strong evidence for the existence of \M\ and the nonexistence of a competent financial regulator.  We are not maligning the public servants who are doing their best in an impossible role.  A few thousand lawyers regulating a large swath of the United States financial industry works about as well in practice as you might expect, and it isn't going to work better and better as finance becomes increasingly complex.}.  If no \M\ exists, then the explanation in \RefsTheExplanation\ is wrong.  Finding \M\ and knowing \M's trading, a regulator should be able to generally match \M's positions and trading with the most strikingly consistent of the rich patterns in \FigUSCompanies.  If not, then the explanation in \RefsTheExplanation\ is wrong~\footnote{Given the track records of financial regulators (Madoff, Wirecard, 2008, etc.), any conclusion a regulator publicly reaches on this issue should be judged on the merits of whatever supporting evidence they provide.  A definitive answer to the question of whose trading caused the extraordinary plots in \FigMain\ would be ideal.  Evidence strongly disfavoring the explanation in \RefsTheExplanation\ would be sufficient.}.

\FigChinaCompanies\ shows what happens when \M\ can't easily expand and contract his long positions.  As insightfully pointed out in Ref.~\cite{qiao2020overnight}, China is the only country that prohibits buying a stock and then selling it later the same day.  In the presence of this restriction in China, the story in \RefsTheExplanation\ predicts no overnight (blue) lines going up or intraday (green) lines going down in any stock trading on an exchange in China~\footnote{There is actually some wriggle room here.  For example, \M\ can expand and contract his long positions over a two day period (expand this morning, contract tomorrow afternoon) rather than daily (expand this morning, contract this afternoon), which is illegal in China.  Alternatively, and more in keeping with the cynical spirit of this article, the mere fact that something is illegal doesn't mean nobody does it.}.  This strong prediction is borne out in all but two of the fifty striking plots in \FigChinaCompanies.  We do not know the cause of the two exceptions (SAIC Motor and Haier Smart Home), but we imagine a highly entertaining story that will never be told in English and ends with someone disappearing.

We have previously noted~\cite{knuteson2016,knuteson2018} that \M\ can stumble upon the Strategy without specifying exactly how that stumbling occurs.  This stumbling upon might happen for a reason an onlooker could guess, but it is far more likely to happen for a reason an insider debugging the code can easily find but an onlooker would never guess.  (Anyone who codes knows the havoc you can wreak with errors as simple as passing data in the wrong units~\cite{MarsClimateOrbiter} or failing to initialize the values of an array to zero.  If you genuinely want to know the cause of some strange behavior, you can straightforwardly determine it by wading into the code and persistently debugging it.  Speculating from afar is usually a waste of time.)  If there is more than one such \M, this stumbling upon could have happened in different ways in each firm.  Importantly, any firm that happened upon the Strategy would have been more likely than others to do well and survive.

Similarly, the reason \M's trading has driven overall prices upward (\FigIndices) might be for a reason an onlooker could guess~\footnote{For example, an onlooker might note that \M\ is generally at liberty to expand his long positions but needs to find shares to borrow to expand his short positions.  This could have led \M\ to start to expand his long positions earlier in the day than his short positions.}, but it is far more likely to be for a reason an insider debugging the code can easily find but an onlooker would never guess.  As in the preceding paragraph, there is a Darwinian-like pressure at work:  any particularly impactful change to \M's trading that tends to obviously and systematically push prices down (over a time period long enough to draw suspicion) is more likely to be flagged, investigated, and corrected (by any self-respectingly paranoid \M\ himself) than a feature of \M's trading that tends to obviously and systematically push prices up (which nobody will question and everybody will be very happy about).

Although we have described an initially accidental \M, quants are not stupid, and any initially accidental \M\ will eventually realize what he is doing~\footnote{If an accidental \M\ has outside investors, then upon discovering his use of the Strategy \M\ must (stop and) inform his current investors of his use of the Strategy~\cite{knuteson2018}.  \M\ must also disclose previous use of the Strategy to any potential future investor.  Failing to inform \M's (current and potential future) investors constitutes a material misrepresentation of how \M\ has made money.  \M's ill-gotten gains in this case include all of \M's profits made with the money of investors (including all public pension funds) who would have withdrawn their money (or not invested) if \M\ had properly disclosed the materiality of his previous accidental use of the Strategy.  The individuals within \M\ who should have made this disclosure and chose not to have committed fraud.  They can and should be criminally charged.}.  From that point on \M\ is malicious, not accidental, and \M's use of the Strategy and \M's ensuring that overall prices go up, not down, is deliberate, not an accident.

The profits \M\ creates for himself using the Strategy are ill-gotten gains.  Any profits \M\ obtains from other bets (including in other asset classes) that are materially aided by \M's systematically pushing up stock prices are also ill-gotten gains~\footnote{Such bets could include, among other things, bets (in various forms) that certain parties will not default, bets (in various forms) that certain interest rate spreads will not widen, and bets (using futures contracts and options, for example) that prices of stock market indices and other correlated assets will rise (or not fall).}.  More than all~\footnote{If \M\ makes one billion dollars with the Strategy and loses a few hundred million dollars on unrelated bets, then \M's ill-gotten gains (one billion dollars) exceed \M's total profits.} of \M's profits may be ill-gotten even if \M\ is a multi-strategy firm and only one component of \M\ employs the Strategy.

So, to recap, the explanation in \RefsTheExplanation\ fits the general pattern of overnight and intraday returns in the United States in \FigUSCompanies, fits China as the exception in \FigIndices\ and the strikingly homogeneous pattern among China's companies in \FigChinaCompanies, handles the negative intraday returns in the other countries in \FigIndices, explains the striking consistency in the blue and green curves in \FigIndices, and predicts (and would explain the existence of and success of) one or more large, long-lived quant firms who have employed the Strategy, for which there are several candidates.

Increasing reliance on automation in all areas of our lives makes it ever more important that we be capable of identifying when an algorithm has gone off the rails and be able to step in, determine what is going on, and take whatever actions are necessary to correct it.  The stunning failure of regulators, economists, and others to recognize \FigMain\ as a problem and take corrective action in an arena as public and closely followed as the world's stock markets is not promising in this regard.

The mixup of units that jeopardized the Mars Climate Orbiter over two decades ago need not have doomed it~\cite{MarsClimateOrbiter}.  From Ref.~\cite{MarsClimateOrbiterIEEESpectrum}:  ``[G]round controllers ignored a string of indications that something was seriously wrong with the craft's trajectory, over a period of weeks if not months.  [And] managers demanded that worriers and doubters `prove something was wrong,' even though classic and fundamental principles of mission safety should have demanded that they themselves, in the presence of significant doubts, properly `prove all is right' with the flight.''

\begin{figure}[thb]
\includegraphics[width=3.45in]{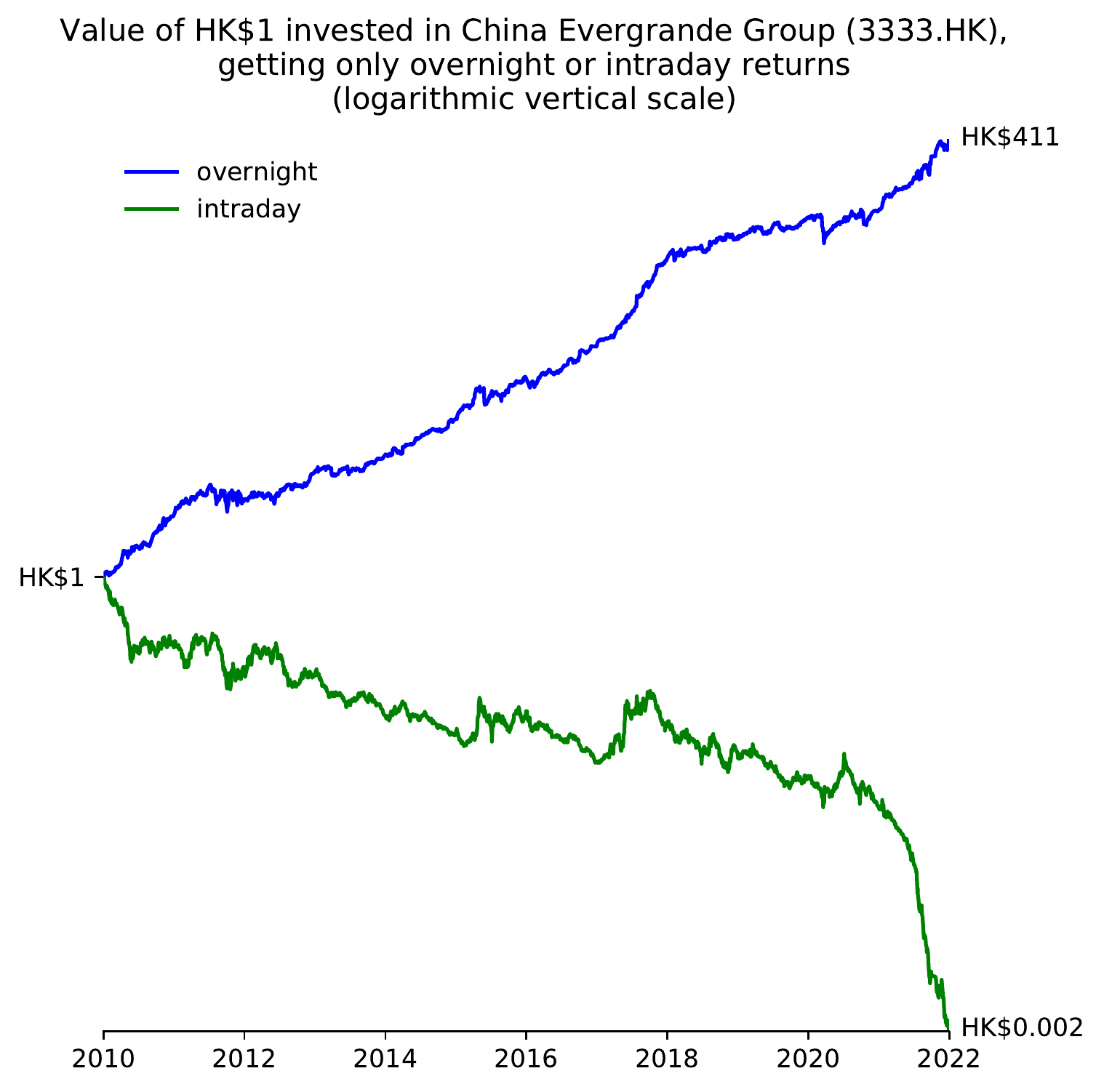}
\caption{\label{fig:Evergrande}The value of one Hong Kong dollar of Evergrande stock, invested at the start of 2010, getting only overnight or intraday returns.  The vertical axis has logarithmic scale.  HK\$1 invested in Evergrande at the start of 2010, getting only intraday returns, would leave you with \ldds{\$0.002} on \lastDataDate, for a cumulative return of \ldds{$-99.8\%$}.}
\end{figure}

Figure~\ref{fig:Evergrande} shows overnight and intraday returns to Evergrande as a teaser, much like we showed \FigMeme\ in Ref.~\cite{knuteson2021}.  If a suspicious return pattern seems to consistently pop up in matters that are of concern for other reasons, you probably want to figure out what is going on.  When something suspicious like this appears, you definitely don't want people who can't figure out what is going on, who likely caused what is going on, or who didn't tell you about what was going on anywhere near your money.

\section{They Still Haven't Told You}

Even if \FigMain\ somehow miraculously turns out to be fine, the failure of others to bring the problem suggested by \FigMain\ to your attention is utterly inexcusable.

The fundamental problem here is not quants, economists, regulators, journalists, and others failing to understand that \FigMain\ is suspicious or that the explanation in \RefsTheExplanation\ is the only one that fits.  The fundamental problem is that these people dislike the implications.  If the world's stock markets had been going down with the consistency they have been going up, these same people would have been all over any suspicious return pattern that might have provided a clue as to why.  

This issue is not rocket science or nuclear physics.  The intellectual skills required are the ability to distinguish lines that go up from lines that go down, and the level of ethical behavior required is a simple willingness to alert others to a glaringly obvious problem.  These are the lowest of low bars.  Remarkably, of the many thousands of people who could have clearly and persistently brought this matter to your attention, we are the only ones who have chosen to do so.  

Every single person who could have warned you about the problem so obvious in \FigMain\ and chose not to has made a significant contribution to the course of human events in the same way so many others have contributed to the greatest train wrecks of history -- by staying silent.  At the time of this writing, their callously self-interested, intellectually dishonest, and cowardly contribution continues.  As if to deliberately emphasize just how little they care about your well-being and your having the information you need to make good decisions, they still haven't told you.

\bibliography{still}

\end{document}